\documentclass[sn-aps]{sn-jnl}% American Physical Society (APS) Reference Style
\usepackage{graphicx}%
\usepackage{multirow}%
\usepackage{amsmath,amssymb,amsfonts}%
\usepackage{amsthm}%
\usepackage{mathrsfs}%
\usepackage[title]{appendix}%
\usepackage{xcolor}%
\usepackage{textcomp}%
\usepackage{manyfoot}%
\usepackage{booktabs}%
\usepackage{algorithm}%
\usepackage{algorithmicx}%
\usepackage{algpseudocode}%
\usepackage{listings}%
\usepackage{array}%
\usepackage{caption}%
\usepackage{upgreek}
\usepackage{siunitx}
\usepackage{comment}
\newcolumntype{P}[1]{>{\centering\arraybackslash}p{#1}}
\usepackage{ulem}

\definecolor{mycolor}{rgb}{0,0,0} %black

\theoremstyle{thmstyleone}%
\theoremstyle{thmstyletwo}%

\theoremstyle{thmstylethree}%
\newcolumntype{R}[1]{>{\raggedleft\arraybackslash}p{#1}} % Right aligned with fixed width

\begin{document}

\title[Article Title]{
Growth of uniform helium submonolayers adsorbed on single-surface graphite observed by surface X-ray diffraction
%Structural relaxation of helium submonolayer adsorbed on single-surface graphite observed by surface X-ray diffraction
%Sample preparation criteria for observing helium submonolayer adsorbed on single-surface graphite by surface X-ray diffraction
}

%%=============================================================%%
%% GivenName	-> \fnm{Joergen W.}
%% Particle	-> \spfx{van der} -> surname prefix
%% FamilyName	-> \sur{Ploeg}
%% Suffix	-> \sfx{IV}
%% \author*[1,2]{\fnm{Joergen W.} \spfx{van der} \sur{Ploeg} 
%%  \sfx{IV}}\email{iauthor@gmail.com}
%%=============================================================%%

\author*[1]{\fnm{Atsuki} \sur{Kumashita}}\email{rk23y004@guh.u-hyogo.ac.jp}
\author*[2]{\fnm{Hiroo} \sur{Tajiri}}\email{tajiri@spring8.or.jp}
%\equalocate{These authors contributed equally to this work.}
\author[3,4]{\fnm{Jun} \sur{Usami}}
%\author[3]{\fnm{Jun} \sur{Usami}}
\author[1]{\fnm{Yu} \sur{Yamane}}
\author[1]{\fnm{Shigeki} \sur{Miyasaka}}
\author[4]{\fnm{Hiroshi} \sur{Fukuyama}}
\author*[1]{\fnm{Akira} \sur{Yamaguchi}}\email{yamagu@sci.u-hyogo.ac.jp}

\affil*[1]{\orgdiv{Graduate School of Science}, \orgname{University of Hyogo}, \orgaddress{\street{3-2-1 Kouto, Kamigori-cho}, \city{Ako-gun}, \postcode{678-1297}, \state{Hyogo}, \country{Japan}}}
\affil*[2]{\orgname{Japan Synchrotron Radiation Research Institute}, \orgaddress{\street{1-1-1 Kouto}, \city{Sayo}, \postcode{679-5198}, \state{Hyogo}, \country{Japan}}}
\affil[3]{\orgdiv{Electronics and Photonics Research Institute}, \orgname{National Institute of Advanced Industrial Science and Technology}, \orgaddress{\street{Central 4-1, 1-1-1 Higashi}, \city{Tsukuba}, \postcode{305-8565}, \state{Ibaraki}, \country{Japan}}}
\affil[4]{\orgdiv{Cryogenic Research Center},\orgname{The University of Tokyo} \orgaddress{\street{2-11-16 Yayoi}, \city{Bunkyo-ku}, \postcode{113-0032}, \state{Tokyo}, \country{Japan}}}
%
%
%%==================================%%
%% Sample for unstructured abstract %%
%%==================================%%
%
\abstract{
%-------------------------------
%
We observed surface X-ray diffraction from $^{4}$He submonolayers adsorbed on a single-surface graphite
using synchrotron X-rays.
Time evolutions of scattering intensities along the crystal truncation rod (CTR) were observed
even after reaching the base low temperature in a selected condition of sample preparation.
Our simulations for CTR scatterings based on the random double-layer model,
in which helium atoms are distributed randomly in the first and second layers with a certain occupancy ratio,
can consistently explain the observed intensity changes.
These results support the scenario that He atoms are stratified initially as a nonequilibrium state and then relaxed into a uniform monolayer by surface diffusion, 
where the relaxation process was observed as a decrease in CTR scattering intensity.
The observed time constant was, however, much longer than those estimated from quantum and thermal surface diffusions.
This implies homogeneous processes in surface diffusions were strongly suppressed by local potentials in such as atomic steps or microcrystalline boundaries.
%------------------------
%
}
% end of abstract

\keywords{Helium, Low dimensional system, surface diffraction, Synchrotron X-rays}
%
%%\pacs[JEL Classification]{D8, H51}
%
%%\pacs[MSC Classification]{35A01, 65L10, 65L12, 65L20, 65L70}
%
\maketitle
%%%%%%%%%%%%%%%%%%%%%%%%%%%%%%%%%%%%%
\section{Introduction}\label{sec1}
Helium (He) films physically adsorbed on graphite are an ideal two-dimensional quantum system~\cite{godfrin1995experimental,fukuyama2008nuclear}
where a variety of two-dimensional quantum phases appear at low temperatures depending on areal densities.
%----------------------------------
%
%
Exfoliated graphite including Grafoil, Papyex, and ZYX have been used as a substrate to measure
heat capacity~\cite{bretz1973phases,elgin1974thermodynamic,bretz1977ordered,greywall1993heat},
NMR~\cite{rapp1993nuclear}, torsional oscillators~\cite{nyeki2017intertwined,choi2021spatially},
and neutron diffraction~\cite{lauter1987neutron,Lauter1991}, 
because they have a large specific-surface area of the order of 20~m$^2$/g.
%,wiechert1991ordering
%
Sample preparation conditions for uniform He films on the exfoliated graphite at temperatures below 4~K
have been well established.
For surface X-ray diffraction (SXRD)~\cite{tajiri2020progress}, however, a single-crystal or optically-flat mosaic crystal with a mosaic spread less than 0.1$^\circ$ is required instead of Grafoil which has a much larger mosaic spread of around 30$^\circ$~\cite{taub1977neutron,takayoshi2010determination}.
There remains a challenge when He films are prepared on a single-surface substrate~\cite{elgin1974thermodynamic,campbell1985he,chae1989microcalorimetry,godfrin1995experimental}.
Eventually, Campbell and Bretz~\cite{campbell1985he} and Chae and Bretz~\cite{chae1989microcalorimetry} reported non-reproducible behaviors depending on adsorption conditions.
\par
%----------------------------------
%In particular, at lower densities than that for the first layer completion, 12.0~nm$^{-2}$~\cite{greywall1993heat},
%the time constant until adsorption equilibrium is extremely long~\cite{godfrin1995experimental}.
%
%----------------------------------
Previously, we reported SXRD from submonolayer He films on a single-surface graphite, 
using a surface buffer consisting of Grafoil to estimate the areal density~\cite{Yamaguchi2022,kumashita2023simulations}.
%---------------------
It should be noted that the effective surface area of a single-surface graphite is extremely small compared to exfoliated graphite.
An adsorbate amount to prepare an He monolayer on a single-surface substrate with a surface area of 1~cm$^{2}$
is approximately 10$^{-8}$~mol, corresponding to a volume of 10$^{-5}$~cm$^3$ at standard temperature and pressure.
Controlling such a small amount of gas is quite challenging.
%===========
Instead, by regulating the adsorbate amount governed by the surface-buffer cell with a large surface area,
we can control the areal densities on a single-surface graphite.
%===========
%--------------------------------------------------------------
He adsorption experiments with small surface areas
have been reported in AC calorimetry of $^{4}$He on graphite~\cite{campbell1985he,chae1989microcalorimetry}
and mechanical resonance of carbon nanotubes with adsorbed $^{3}$He~\cite{todoshchenko2022topologically,todoshchenko2024quantum}
using a surface buffer.
\par
%----------------------------------
%
%
In this article, we report SXRD from $^{4}$He submonolayers adsorbed on a single-surface graphite using synchrotron X-rays.
Time evolutions of crystal truncation rod (CTR) scattering intensities were observed even when reaching the base temperature under careful control of sample temperature and vapor pressure.
Our simulation results using the random double-layer model consistently support the scenario that
stratified He layers in nonequilibrium states relaxed to a uniform monolayer by surface atomic diffusion and that the relaxation process was observed
as a decrease in CTR scattering intensity.
%
%=================================
%%%%%%%%%%%%%%%%%%%%%%%%%%%%%%%%%%%%%

\section{Materials and Methods}\label{sec2}
%===========================================
%
% SXRD
%
%-------------------------------------------
SXRD were performed at the beamline BL29XU~\cite{tamasaku2001spring} in SPring-8.
Using synchrotron X-rays of 30~keV in energy,
we measured CTR scatterings along the $00L$ rod~\cite{robinson1986crystal,tajiri2020progress}. 
%--------------------------------------------------
Each measurement time over $L$ from 0.1 to 2.0 was 10~min at $L$ intervals of 0.1 with an exposure time of 20~s for each. 
The $00L$ CTR scatterings provide structural information about the distributions of surface atoms perpendicular to a surface.
%%------------------------------------

%------------------------------------
A highly oriented pyrolytic graphite (HOPG) having a narrow mosaic spread less than 0.1$^\circ$~\cite{Tajiri_unpublished}
was used as a substrate for SXRD. 
%-----------------------------
To ensure thermal contact,
a glass of 10 mm (length) $\times$ 5 mm (width) $\times$ 1.7 mm (height), on which a 50~$\upmu$m thick HOPG was grown, was adhered to a Cu base using Ag pastes, which were \textcolor{mycolor}{heat cured} to the edges of the HOPG, sides, and bottom of the glass.
A W-wire heater (25~ohm) was attached using ceramic adhesive onto the \textcolor{mycolor}{heat-cured} Ag pastes
at one side of the HOPG. 
The substrate including the Cu base was pre-baked at 150$^{\circ}$C in a high vacuum of the order of 10$^{-5}$~Pa for 24~h,
and then installed into the sample cell of the cryogenic system
in the ultra-high vacuum SXRD instrument.

%===========================================
%
% Instrumentation and sample preparation
%
%-------------------------------------------
Our He-inlet line consisted of a room-temperature gas handling system,
a surface-buffer cell containing Grafoil sheets with a large surface area, and a sample cell for X-ray observation.
These components were connected each other by capillaries of inner diameter of 0.6~mm.
%---------------------------
The buffer cell was prepared as follows:
Grafoil disks of GTA grade, which were diffusion bonded with Ag foils, were
installed in a Cu cylinder and sealed with Stycast 1266. 
The surface area of the Grafoil disks was determined to be 5.38~m$^2$
by N$_{2}$ adsorption isotherm at 77~K.
%------------------------
When the vapor pressures of $^4$He films on graphite in the buffer and sample cells are the same in thermal equilibrium,
we can expect their areal densities are also the same.
%---------------------------
Other details of the sample cell, cryogenic system with a GM refrigerator and a 1K-pot, and SXRD instrument
have been described in the previous publications~\cite{Yamaguchi2022,kumashita2023simulations}. 
%
%-----------------------------
%-----------------------------

%============================================
%
% Pretreatment of substrate
%
%-------------------------------------------
The pretreatment of the HOPG substrate in the sample cell was as follows.
To remove impurity gases including moisture,
the whole system was baked at 80$^{\circ}$C for 48~h, and
the He-inlet line was flushed several times by He gas at room temperature.
Following these treatments, pressure in the sample cell was maintained less than 1$\times$10$^{-4}$~Pa at room temperature. 
During SXRD experiments at low temperatures, the substrate was degassed using the W-wire heater at 100, 50, and 2.8~K, respectively.

%===========================================
%
% Sample preparation
%
%-------------------------------------------
He sample gas, of which amount was measured in the gas handling system,
was introduced through the capillary to the sample and buffer cells.
Then, the substrate was annealed
to obtain uniform He films, using a resistive heater attached to 1K-pot.
%~\cite{elgin1974thermodynamic,campbell1985he,chae1989microcalorimetry,godfrin1995experimental}
%
During the introduction of He gas and annealing the He films,
the pressure, $P_{\text{H}}$, was monitored using a pressure gauge (Setra Sytems Inc., Model 204) in the gas handling system.
%------------------
%
%
The substrate temperature was maintained at $T_{\text{anneal}}$ for a time, $\Delta t_{\text{anneal}}$, 
and then the temperature was reduced down to $T_{\text{L}}$, at which the pressure was $P_{\text{L}}$, at a cooling rate of $dT/dt$. \textcolor{mycolor}{In appropriate $dT/dt$ and $P_{\text{L}}$, both mass imbalance and transport of He between the sample and buffer are expected to be negligible.}
Prior to observing CTR scatterings, we maintained the sample temperature at $T_{\text{L}}$ = 2.8~K for a time, \textcolor{mycolor}{$t_{\text{wait}}$}.
%------------------
$T_{\text{anneal}}$ was 10 -- 14~K for the $\sqrt{3}\times\sqrt{3}$ commensurate solid samples ($\rho ~=~$6.37~nm$^{-2}$)
and 8 -- 13~K for the incommensurate (IC) solid ones (10.6~nm$^{-2}$)~\cite{greywall1993heat}, respectively.
%-----------------
Since it has been reported that time to reach adsorption pressure equilibrium was approximately 10~min
by maintaining the pressure in the range of 0.1 to 1~kPa during annealing~\cite{elgin1974thermodynamic,godfrin1995experimental},
$P_{\text{H}}$ was kept approximately that range.
%

%%%%%%%%%%%%%
%\begin{figure}[t]
%  \centering
%  \includegraphics[width=6cm]{Fig1.eps}
%  \caption{Annealing procedure in preparation of He films on graphite for the present CTR scattering experiments.}
%  \label{fig:anneal}
%\end{figure}
%%%%%%%%%%%%%

%%%%%%%%%%%%%%%%%%%%%%%%%%%%%%%%%%%%%%%%%%%%%%%%%%%%%%%%%%%
\section{Results and Discussion}\label{sec3}
%----------------------------------------------------------------------------------------------------------------------------------
%=================================================
%
% Sample 1: Results and Interpretation 
%
%%%%%%%%%%%%%%%%%%%%%%%%%%%%%%%%%%%%%%%%%%%%%%%%%%%%%%%%%%%%%%%%%%%%%%%%%
\begin{figure}[b]
  \centering
  \includegraphics[width=13cm]{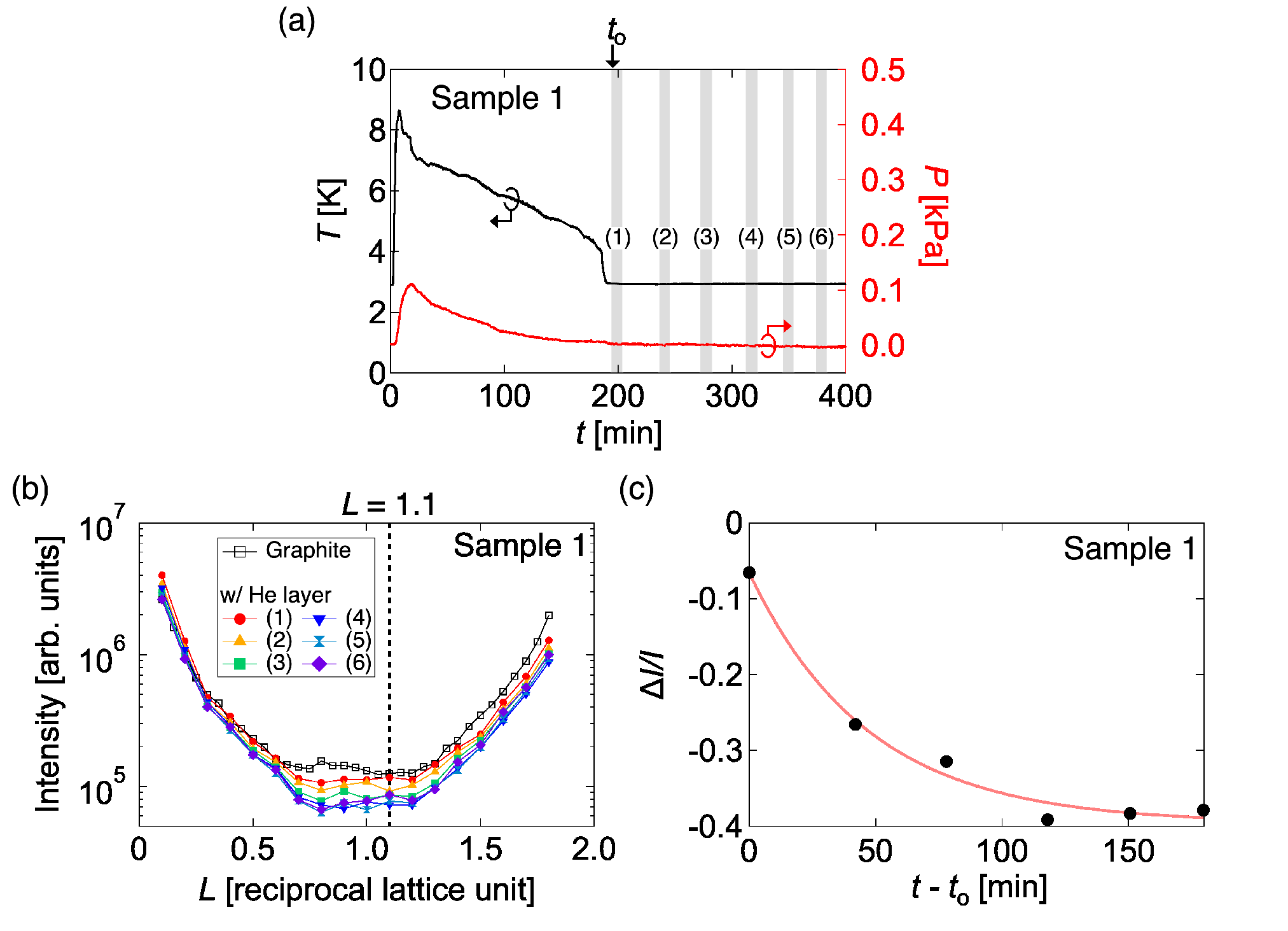}
  \caption{
(a) Time evolutions of temperature and pressure during the adsorption process in Sample~1. 
CTR scattering measurements were performed during the times indicated by the gray bands, (1)--(6). 
(b) CTR scattering intensities observed at the six successive measurements in (a). 
(c) Relative intensity change, $\Delta I/I$, of the CTR scatterings at $L$~=~1.1 
(the vertical dotted line in (b))
plotted as a function of time measured from $t_0$ in (a). 
The red solid line shows the exponential curve that best fits the data. 
}
  \label{fig:CTRrelax}
\end{figure}
%%%%%%%%%%%%%%%%%%%%%%%%%%%%%%%%%%%%%%%%%%%%%%%%%%%%%%%%%%%%%%%%%%%%%%%%%%%%%%
%-------------------------------------------------
%
The time evolutions of temperature and pressure during preparation of the IC phase (Sample 1: 10.6\,nm$^{-2}$) are shown in Fig.~\ref{fig:CTRrelax}(a).
After reaching $T_{\text{L}}$ ($= 2.8$\,K), CTR scattering measurements were made repeatedly at times indicated by the gray bands (1)--(6).
In these measurements, as shown in Fig.~\ref{fig:CTRrelax}~(b), the CTR scattering intensities decreased monotonically with time.
Since the CTR scattering, which is interpreted as amplitude interference
between scattered X-rays from a substrate and surface atoms,
carries structural information at the atomic level~\cite{tajiri2020progress},
these intensity decreases should originate from the structural change of the He sample itself.
Meanwhile, contributions from surface atoms to CTR scatterings are
pronounced at far from Bragg peaks (in this case around $L$ = 1.0, where Bragg reflection is forbidden).
Thus, we plotted the relative change, $\Delta I/I$, at $L$ = 1.1 in the scattering intensity, $I$,
that is measured for the clean graphite, as shown in Fig.~\ref{fig:CTRrelax} (c).
$\Delta I/I$ decreased to approximately $-$0.4,
which is roughly consistent with our simulations assuming a uniform submonolayer He film~\cite{Yamaguchi2022,kumashita2023simulations}. 
By fitting this trend with an exponential function, the time constant was estimated to be approximately 50~min.
%
%---------------------------------------------------------------------------------------------------------------------------------
%
%\textcolor{red}{\sout{Because the intensity \textcolor{red}{decrease was} observed 
%\textcolor{red}{even} after reaching \textcolor{red}{\sout{to}} $T_{\text{L}}$, at which  $P_{\text{L}}$ was less than the detection limit of the \textcolor{red}{pressure gauge},
%it is concluded that the $\Delta I/I$ behavior originated from structural evolutions of the He film at the atomic level 
%from the aforementioned CTR scattering feature.}}
%
%

\begin{figure}[t]
  \centering
  \includegraphics[width=13cm]{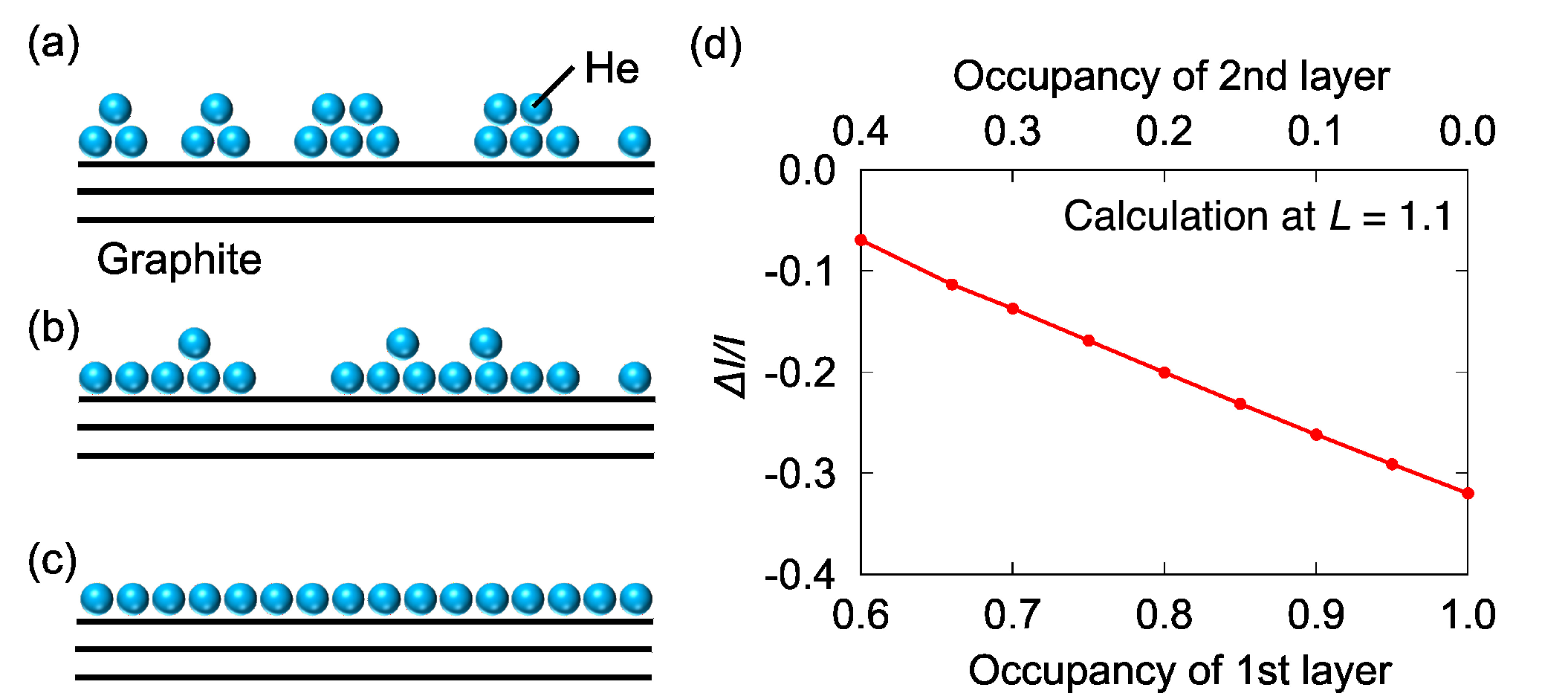}
  \caption{
Schematics in the random distribution model of He adatoms
at the first layer occupancies of (a) 0.66, (b) 0.8, and (c) 1.0.
(d) Calculated intensity change of CTR scatterings at $L$ = 1.1 using the random model
for different first-layer occupancies. 
The total adsorption areal density was fixed at 10.6~nm$^{-2}$.
}
  \label{fig:random}
\end{figure}

We performed simulations using the random double-layer adsorption model
based on the oblique bilayer structure~\cite{Carneiro1981,wiechert1991ordering}.
In this model, He atoms are randomly distributed forming single or double layer
with the fixed number of atoms corresponding to a uniform IC phase of 10.6~nm$^{-2}$ (see Figs.~\ref{fig:random} (a) -- (c)).
Figure~\ref{fig:random}(d) shows the simulation result of $\Delta I/I$ at $L$ = 1.1 for the IC solid phase,
exhibiting almost linear behavior with the occupancy of the first layer.
%--------------------------------------------------
%
Thus, we speculate that the observed change in $\Delta I/I$ (Fig.~\ref{fig:CTRrelax}(c))
corresponds to relaxation of inhomogeneously distributed bilayer films into a uniform monolayer.
%============================================

%-----------------------------------------------------------------------------------------------------------------------
%============================================
%
% Consideration on surface diffusion
%
%--------------------------------------------
%%%%%%%%%%%%%%%%%%%%%%%%%%%%%%%%%%%%%%%%%%%%%%%%%%%%%%%%%%%%%%%%%
%%%%%%%%%%%%%%%%%%%%%%%%%%%%%%%%%%%%%%%%%%%%%%%%%%%%%%%%%%%%%%%%%%
%
% Quantum to Thermal process
%
%============================================
%%%%%%%%%%%%%%%%%%%%%%%%%%%%%%%%%%%%%%%%%%%%%%%%%%%%%%%%%%%%%%%%%%
%
%
%To confirm this scenario, we estimated the stratification relaxation time due to surface diffusion of He
To confirm this scenario, we estimated the demotion relaxation time due to surface diffusion of He
via quantum tunneling, at first.
Assuming a rectangular potential barrier of a width, $l_0$, and height, $E_\text{d}$,
the surface diffusion coefficient, $D_{\text{q}}$, of a particle of the mass, $m$, is represented
within the WKB approximation~\cite{johnston1966gas} as follows:
%%%%%%%%%%%%%%%%%%%%%%%%%%%%%%%%%%%%%%%%%%%%%%%%%%%%%%%%%%%%%%%%%%%%
\begin{align}
        D_{\text{q}} = \frac{1}{2} a^{2} \nu_{\parallel} \exp \left[ -\frac{2 l_{0} \left( 2mE_{\text{d}} \right)^{\frac{1}{2}}}{\hbar} \right],
    \label{eq:diffusion}
\end{align}
%%%%%%%%%%%%%%%%%%%%%%%%%%%%%%%%%%%%%%%%%%%%%%%%%%%%%%%%%%%%%%%%%%%%
%
where $a$, $\nu_\parallel$, and $\hbar$
represent lattice periodicity,
vibrational frequency in the in-plane direction,
and reduced Planck's constant, respectively.
%----------------------
$E_{\text{d}}$ (or the diffusion energy in Ref.~\cite{johnston1966gas}) of He atom in the second layer was estimated to be 5~K
from the calculation of the potential corrugation in the second layer~\cite{whitlock1998monte}.
%
%\textcolor{blue}{\sout{Since the densities of second layer is dilute, the interactions between adsorbed atoms in the second layer can be neglected. 
%
%The values of $a$ and $l_0$ can be estimated based on the corrugation potential in the first-layer IC solid.}(書くならthe simplest caseとしてneglecting interaction between the 2nd layer atoms程度で簡単に)}
%
We adopted the nearest-neighbor distances in the first-layer IC solid of He (3.15~$\text{\AA}$~\cite{Carneiro1981}) as $a$,
while $l_{0}$ was set to 2~$\text{\AA}$ according to the corrugation potential periodicity in the first-layer IC solid.
%
%----------------------
Substituting these values to  Eq.~(\ref{eq:diffusion}) and assuming $\nu_\parallel$ of the order of $10^{11} \, \text{Hz}$,
$D_{\text{q}}$ was estimated to be approximately $10^{-11} \, \text{m}^2/\text{sec}$.
This value is in the same order as that reported for HD and D$_{2}$ on graphite~\cite{Wiechert2003physics}.
The relaxation time for the demotion, $\tau$, of this system
would be expressed as $\tau = L_{\text{c}}^{2}/4D_{\text{q}}$ in random surface diffusion
with a crystalline size of graphite, $L_{\text{c}}$~\cite{van1993theoretical}.
%----------------------
Assuming $L_{\text{c}}$ = 10~$\upmu$m for the HOPG substrate,
$\tau$ is estimated to be a few tenths of seconds, which is much shorter than the observed time constant (50 min)
in the present experiment.

%---------------------------------------
Next, we considered a thermal process at $T_{\text{L}}$ (2.8 K)
using the thermally activated diffusion coefficient, $D_{\text{t}}$, defined as~\cite{van1993theoretical}
%%%%%%%%%%%%%%%%%%%%%%%%%%%%%%
% Thermal diffusionの式
%~\cite{van1993theoretical}. 
\begin{align}
    D_{\text{t}} = \frac{1}{2} a^2 \nu_{\parallel} \exp\left(-\frac{E_\text{d}}{k_{\text{B}} T_{\text{L}}}\right),
    \label{eq:thermal}
\end{align}
%%%%%%%%%%%%%%%%%%%%%%%%%%%%%%
where $k_{\text{B}}$ denotes Boltzmann constant.
With the parameters used for quantum diffusion,
this process also provided a much shorter diffusion time constant (0.03~s) than the observation.

The fact that the obtained relaxation time (50~min) was much longer than those estimated both for the thermal and quantum surface diffusions
implies that these homogeneous surface diffusion processes were strongly suppressed.
%----------------------
For example, localizations at atomic steps on a graphite surface or microcrystalline boundaries, which have large and heterogeneous adsorption potentials,
would substantially elongate the relaxation time compared to those estimated from the elementary processes described above.
%----------------------
We note that similarly long relaxation times exceeding 1~h have been previously reported for He films adsorbed on graphite
with adsorption isotherms~\cite{godfrin1995experimental} and manometry~\cite{elgin1974thermodynamic}.

%============================================

\begin{figure}[ht]
  \centering
  \includegraphics[width=13cm]{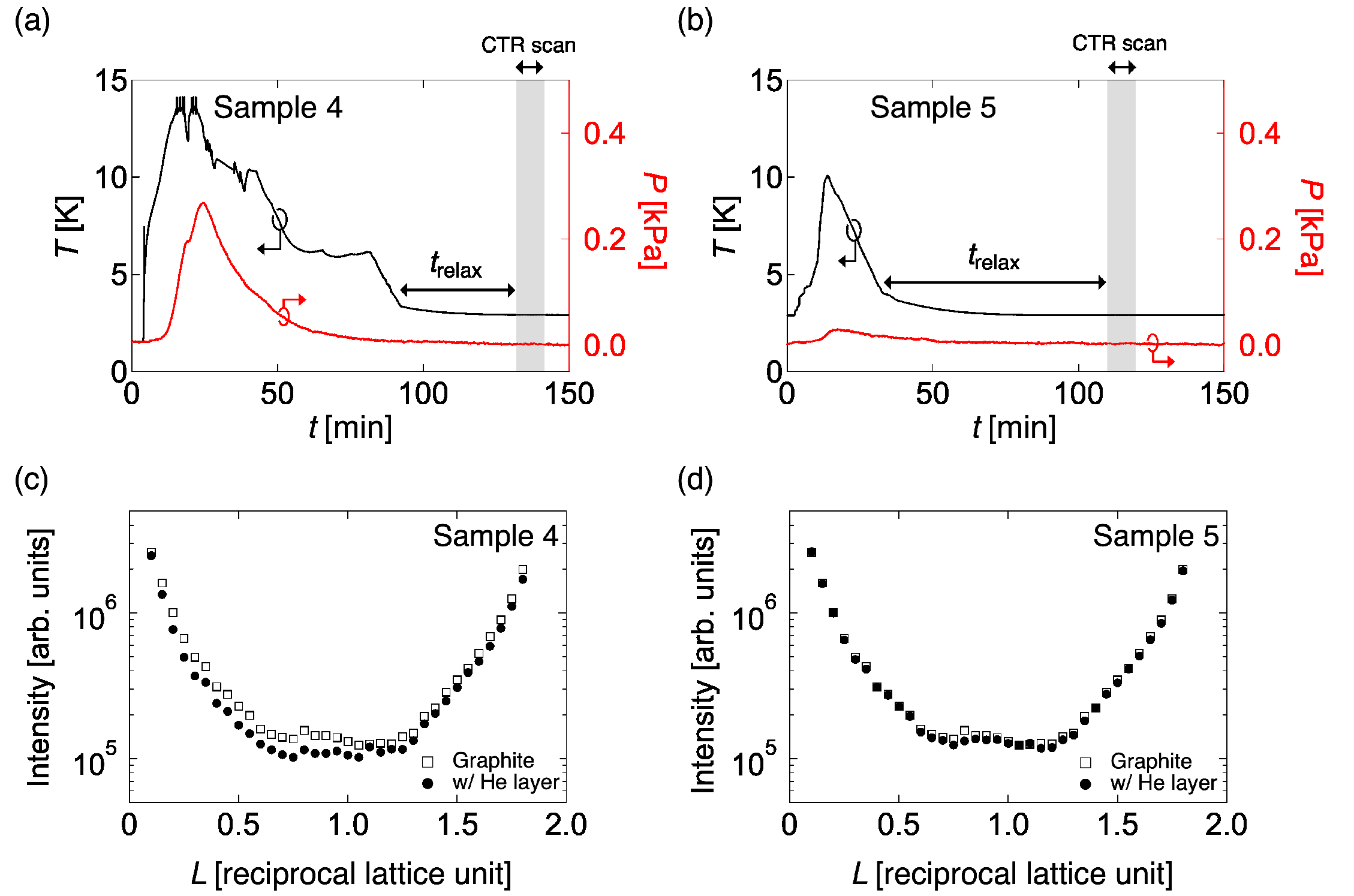}
  \caption{
Time evolutions of temperature and pressure during annealing in (a) Sample~4 and (b) Sample~5. 
CTR scattering intensities (closed circle) observed after the annealing are plotted
in (c) and (d) for Sample~4 and Sample~5, respectively. 
Results for clean graphite (open square) are also plotted for comparison. 
Statistical errors are smaller than the size of the plotted symbols.
}
  \label{fig:CTRexample}
\end{figure}

%===========================================
%
% Results for obs. 4 and 5 and the others.
%
%-------------------------------------------
%%%%%%%%%%%%%%%%%%%%%%%%%%%%%%%%%%%%%%%%%%%%%%%%%%%%%%%%%%%%%%%%%%%%%%%%%
%
% Discussion on sample preparations
%
For further information,
table~\ref{tab:obs} lists the sample preparation conditions employed for all samples prepared in the present CTR scattering measurements including those of the C phase (6.37\,nm$^{-2}$). 
Here, if the relaxation of the CTR scattering intensity was observed (Yes) or not (No) is indicated at the last column.
In addition to Sample 1, we observed the CTR scattering intensity change in Sample 4.
For comparison, the time evolutions of temperature and pressure recorded for Samples 4 and 5 as well as the obtained scattering profiles,
are shown in Fig.~\ref{fig:CTRexample}.
%-------------------------
Regarding Sample~4, while temperature control was not smooth, 
the sample was heated to 13~K and then maintained at 6 K for 0.5~h until pressure equilibrium was reached,
after which it was slowly cooled to the base temperature (see Fig.~\ref{fig:CTRexample}(a)).
%======================================
% Disucssion fo PL
Conversely, in Samples 2, 3, and 6, where $P_{\text{L}}$ values were rather high,
no clear CTR scattering intensity changes were observed.
%
%This may indicate too much density inhomogeneities left in these He films, {\color{magenta}which prevented us from observing the smooth time evolution of the CTR scattering intensity.}
%
This may indicate too much density inhomogeneities left in these He films,
which prevented us from observing a distinct time evolution of the CTR scattering intensity.
%-------------------------
%
Besides, in Sample~5, $P_{\text{H}}$ was considerably lower than that in Sample 4,
%which suggests mass transport on a single-surface HOPG was not enough.  \textcolor{magenta}{$\leftarrow$OR (suggests that mass transport between the sample and buffer cells or over HOPG crystalline boundaries in the sample cell was suppressed due to the lack of sufficient vapor pressure)}
which suggests that mass transport between the sample and buffer cells or over HOPG crystalline boundaries in the sample cell was suppressed due to the lack of sufficient vapor pressure.
From these experiences,
we infer that maintaining sufficient pressure and equilibrium during annealing is essential to form well-defined uniform He submonolayer films.
This criterion is consistent with the one previously noted for Grafoil~\cite{elgin1974thermodynamic,godfrin1995experimental}. 
%=================================================

%%%%%%%%%%%%%%%%%%%%%%%%%%%%%%%%%%%%%%%%%%%%%%%%%
\begin{table}[t]
\centering
\caption{Sample preparation conditions of $^4$He films on graphite for CTR scattering experiments.}
\setlength{\tabcolsep}{4pt}% default is 6pt
\begin{tabular}{P{1.0cm}P{1.0cm}R{0.5cm}R{0.5cm}R{0.8cm}R{0.8cm}R{0.5cm}R{0.8cm}R{1.0cm}P{1.4cm}} % Use P{width} for the last column
\hline
\hline
\multicolumn{1}{c}{Sample} &
\multicolumn{1}{c}{Phase} & 
\multicolumn{1}{c}{$\rho$} &
\multicolumn{1}{c}{$P_\text{H}$} &
\multicolumn{1}{c}{$T_\text{anneal}$} &
\multicolumn{1}{c}{$\Delta t_\text{anneal}$} &
\multicolumn{1}{c}{$\frac{dT}{dt}$} &
\multicolumn{1}{c}{$P_\text{L}$} &
\multicolumn{1}{c}{\textcolor{mycolor}{$t_\text{wait}$}} &
\multicolumn{1}{c}{CTR scattering} \\
\multicolumn{1}{c}{No.} & & \multicolumn{1}{c}{$[\text{nm}^{-2}]$} & \multicolumn{1}{c}{[kPa]} & \multicolumn{1}{c}{[K]} & \multicolumn{1}{c}{[h]} & \multicolumn{1}{c}{[K/h]} & \multicolumn{1}{c}{[kPa]} & \multicolumn{1}{c}{[h]} & \multicolumn{1}{c}{intensity change}\\
\hline
1 & IC &  10.6  & 0.10  & 8  & $<$\!\! 0.01 & 1  & $<$\!\! 0.01 & 0.01 \!--\! 3    & \textcolor{red}{Yes} \\
2 &    &  10.6  & 0.70  & 13 & $<$\!\! 0.01 & 3  & 0.15         & $<$\!\! 0.01     & \textcolor{blue}{No} \\
3 &    &  10.6  & 0.14  & 8  & 0.17         & 3  & 0.02         & $<$\!\! 0.01     & \textcolor{blue}{No} \\
4 & $\sqrt{3}\!\times\!\!\sqrt{3}$ & ~~6.37 & 0.25 & 12 \!--\! 14 & 0.17 & 12 &      $<$\!\! 0.01 & 0.7\,\,\, & \textcolor{red}{Yes} \\
5 &    & ~~6.37 & 0.025 & 10 & $<$\!\! 0.01 & 12 & $<$\!\! 0.01 & 1.0\,\,\,        & \textcolor{blue}{No} \\
6 &    & ~~6.37 & 0.095 & 12 & $<$\!\! 0.01 & 2  & 0.02         & $<$\!\! 0.01     & \textcolor{blue}{No} \\
\hline
\hline
\end{tabular}
\label{tab:obs}
\end{table}
%%%%%%%%%%%%%%%%%%%%%%%%%%%%%%%%%%%%%%%%%%%%%%%%%%%%%%%%%%%%%%%%%%%%%%%%%%
%%%%%%%%%%%%%%%%%%%%%%%%%%%%%%%%%%%%%%%%%%%%%%%%%%%%%%%%%%%%%%%%%%%%%%%%%
%%%%%%%%%%%%%%%%%%%%%%%%%%%%%%%%%%%%%%%%%%%%%%%%%%%%%%%%%%%%%%%%%%%%%%%%%

\section{Conclusion}
We observed SXRD from $^{4}$He submonolayers adsorbed on a single-surface graphite
under various $^{4}$He adsorption conditions of temperature and pressure.
Even after reaching the base temperature of 2.8~K, 
CTR scattering intensities continued to evolve over several hours
until converging at the expected submonolayer density.
%{\color{magenta}Even after reaching the base temperature of 2.8~K, 
%CTR scattering intensities continued to evolve over several hours, 
%until converging at the expected submonolayer density.}
%ideal value for a submonolayer He film on graphite.
%------------
Our simulation results using the random double-layer model
revealed that the observed CTR scattering intensity decrease reflects the relaxation process of initially stratified He films in nonequilibrium states
to a uniform monolayer by surface diffusion.
The fact that the observed relaxation time (50\,min) was much longer than those estimated both for thermal and quantum surface diffusions
implies that the homogeneous surface diffusion processes were strongly suppressed by local potentials in such as atomic steps or microcrystalline boundaries.
We also found a proper annealing condition to prepare uniform submonolayer He films on a single-surface graphite substrate.
%%%%%%%%%%%%%%%%%%%%%%%%%%%%%%%%%%%%%
\section*{Acknowledgements}
The synchrotron radiation experiments were performed on RIKEN beamline BL29XU at SPring-8 (proposal Nos. 20220087, 20230040, and 20240032).
This work was partly supported by JSPS KAKENHI (grant No. 22H03883), JST SPRING, Japan (grant No. JPMJSP2175),
and the Sasakawa Scientific Research Grant from The Japan Science Society.
%

%%%%%%%%%%%%%%%%%%%%%%%%%%%%%%%%%%%%%
%%%%%%%%%%%%%%%%%%%%%%%%%%%%%%%%%%%%%
%%%%%%%%%%%%%%%%%%%%%%%%%%%%%%%%%%%%%
%\bibliography{sn-bibliography}

\begin{thebibliography}{10}
\providecommand{\url}[1]{{#1}}
\providecommand{\urlprefix}{URL }
\providecommand{\doi}[1]{\url{https://doi.org/#1}}
\bibcommenthead

\bibitem{godfrin1995experimental}
H.~Godfrin, H.~J. Lauter, in Progress in Low Temperature Physics, vol.~14
  (Elsevier, 1995), p. 213

\bibitem{fukuyama2008nuclear}
H.~Fukuyama, J. Phys. Soc. Jpn. \textbf{77}, 111013 (2008)

\bibitem{bretz1973phases}
M.~Bretz, J.~G. Dash, D.~C. Hickernell, E.~O. McLean, O.~E. Vilches, Phys. Rev. A
  \textbf{8}, 1589 (1973)

\bibitem{elgin1974thermodynamic}
R.~L. Elgin, D.~L. Goodstein, Phys. Rev. A \textbf{9}, 2657 (1974)

\bibitem{bretz1977ordered}
M.~Bretz, Phys. Rev. Lett. \textbf{38}, 501 (1977)

\bibitem{greywall1993heat}
D.~S. Greywall, Phys. Rev. B \textbf{47}, 309 (1993)

\bibitem{rapp1993nuclear}
R.~E. Rapp, H.~Godfrin, Phys. Rev. B \textbf{47}, 12004 (1993)

\bibitem{nyeki2017intertwined}
J.~Ny{\'e}ki, A.~Phillis, A.~Ho, D.~Lee, P.~Coleman, J.~Parpia, B.~Cowan,
  J.~Saunders, Nat. Phys. \textbf{13}, 455 (2017)

\bibitem{choi2021spatially}
J.~Choi, A.~A.~Zadorozhko, J.~Choi, E.~Kim, Phys. Rev. Lett. \textbf{127},
  135301 (2021)

\bibitem{lauter1987neutron}
H.~J.~Lauter, H.~P.~Schildberg, H.~Godfrin, H.~Wiechert, R.~Haensel, Can. J.
  Phys. \textbf{65}, 1435 (1987)

\bibitem{Lauter1991}
H.~J.~Lauter, H.~Godfrin, V.~L.~P.~Frank, P.~Leiderer, Phase Transitions in
  Surface Films 2 (Springer, Boston, 1991), p. 135

\bibitem{tajiri2020progress}
H.~Tajiri, Jpn. J. Appl. Phys. \textbf{59}, 020503 (2020)

\bibitem{taub1977neutron}
H.~Taub, K.~Carneiro, J.K. Kjems, L.~Passell, J.~P.~McTague, Phys. Rev. B
  \textbf{16}, 4551 (1977)

\bibitem{takayoshi2010determination}
S.~Takayoshi, H.~Fukuyama, J. Low Temp. Phys. \textbf{158}, 672 (2010)

\bibitem{campbell1985he}
J.~H.~Campbell, M.~Bretz, Phys. Rev. B \textbf{32}, 2861 (1985)

\bibitem{chae1989microcalorimetry}
H.~B.~Chae, M.~Bretz, J. Low Temp. Phys. \textbf{76}, 199 (1989)

\bibitem{Yamaguchi2022}
A.~Yamaguchi, H.~Tajiri, A.~Kumashita, J.~Usami, Y.~Yamane, A.~Sumiyama,
  M.~Suzuki, T.~Minoguchi, Y.~Sakurai, H.~Fukuyama, J. Low Temp. Phys.
  \textbf{208}, 441 (2022)

\bibitem{kumashita2023simulations}
A.~Kumashita, H.~Tajiri, A.~Yamaguchi, J.~Usami, A.~Sumiyama, Y.~Yamane,
  M.~Suzuki, M.~Minoguchi, Y.~Sakurai, H.~Fukuyama, JPS Conf. Proc.
  \textbf{38}, 011004 (2023)

\bibitem{todoshchenko2022topologically}
I.~Todoshchenko, M.~Kamada, J.P. Kaikkonen, Y.~Liao, A.~Savin, M.~Will,
  E.~Sergeicheva, T.~S.~Abhilash, E.~Kauppinen, P.~J.~Hakonen, Nat. Commun.
  \textbf{13}, 5873 (2022)

\bibitem{todoshchenko2024quantum}
I.~Todoshchenko, M.~Kamada, J.~P.~Kaikkonen, Y.~Liao, A.~Savin, E.~Kauppinen,
  E.~Sergeicheva, P.~J.~Hakonen, Phys. Rev. B \textbf{109}, 224519 (2024)

\bibitem{tamasaku2001spring}
K.~Tamasaku, Y.~Tanaka, M.~Yabashi, H.~Yamazaki, N.~Kawamura, M.~Suzuki,
  T.~Ishikawa, Nucl. Instr. and Meth. A \textbf{467}, 686 (2001)

\bibitem{robinson1986crystal}
I.~K.~Robinson, Phys. Rev. B \textbf{33}, 3830 (1986)

\bibitem{Tajiri_unpublished}
H.~Tajiri, et~al., To be appear

\bibitem{Carneiro1981}
K.~Carneiro, L.~Passell, W.~Thomlinson, H.~Taub, Phys. Rev. B \textbf{24}, 1170
  (1981)

\bibitem{wiechert1991ordering}
H.~Wiechert, Physica B: Condens. Matter \textbf{169}, 144 (1991)

\bibitem{johnston1966gas}
H.~Johnston, Gas Phase Reaction Rate Theory (Ronald Press Company,
  California, 1966), p. 362

\bibitem{whitlock1998monte}
P.~A.~Whitlock, G.~V.~Chester, B.~Krishnamachari, Phys. Rev. B \textbf{58}, 8704
  (1998)

\bibitem{Wiechert2003physics}
H.~Wiechert, in Physics of Covered Solid Surfaces, ed. by A.~P.~Bonzel
  (Springer, Berlin, 2003), pp. 166 -- 299

\bibitem{van1993theoretical}
J.~H. {van der}~Merwe, Interface Sci. \textbf{1}, 77 (1993)

\end{thebibliography}
% common bib file
%%%%%%%%%%%%%%%%%%%%%%%%%%%%%%%%%%%%%
%%%%%%%%%%%%%%%%%%%%%%%%%%%%%%%%%%%%%
\end{document}